\documentclass[a4paper]{jpconf}
\usepackage{graphicx}

\usepackage{SIunits}

\begin{document}
\title{Electronics for the STEREO experiment}

\author{Victor H\'ELAINE on behalf of the STEREO collaboration}

\address{LPSC, CNRS-IN2P3/UGA, 53 Avenue des Martyrs, 38000 Grenoble, FRANCE}

\ead{victor.helaine@lpsc.in2p3.fr}

\begin{abstract}
The STEREO experiment, aiming to probe short baseline neutrino oscillations by precisely measuring reactor anti-neutrino spectrum, is currently under installation. It is located at short distance from the compact research reactor core of the Institut Laue-Langevin, Grenoble, France. Dedicated electronics, hosted in a single \micro TCA crate, were designed for this experiment. In this article, the electronics requirements, architecture and the performances achieved are described. It is shown how intrinsic Pulse Shape Discrimination properties of the liquid scintillator are preserved and how custom adaptable logic is used to improve the muon veto efficiency.
\end{abstract}

\section{Introduction}

The goal of the STEREO experiment is to probe the existence of light sterile neutrinos by precisely measuring the electron anti-neutrino ($\bar{\nu_e}$) energy spectrum at short distance (10\,m) from the compact reactor core of the Institut Laue-Langevin (ILL) (see \cite{Helaine2016,Minotti2016}). The unambiguous signature of this oscillation is to observe a $\bar{\nu_e}$ spectrum distortion dependent on the distance from the core. This will be done using a segmented detector containing Gd-doped liquid scintillator.
The process used to detect $\bar{\nu_e}$ is the Inverse Beta Decay (IBD), depicted in Eq.\;\ref{eq:ibd}. 
\begin{equation}
\bar{\nu_e}+p\rightarrow e^++n
\label{eq:ibd}
\end{equation}
The signature of this process is a prompt signal from the positron energy deposit -- starting from about 2\,MeV -- detected in a delayed coincidence of a few tens of \micro s with the 8\,MeV $\gamma$-cascade produced by the neutron capture on a Gd nucleus.

Heavy shieldings are used to reduce to an acceptable level the high neutron and $\gamma$ background, due to the surrounding neutron beam lines (see more details about background characterisation in \cite{Kandzia2016}). Correlated background is produced by either spallation neutrons from muons or by fast neutrons from the reactor or neighbouring beam lines, with a prompt signal from neutron thermalisation and then a delayed capture signal. Cosmic induced neutrons should be mainly vetoed using a water Cherenkov muon detector located above the detector. In parallel, the liquid scintillator (see \cite{Buck2016}) has been developed to enable Pulse Shape Discrimination (PSD) between signals from $e^+$ and from proton recoils due to fast neutron interactions in the detector.

In order to exploit the full potential of the muon veto and of the PSD for background discrimination,  dedicated electronics has been designed and fully tested for the STEREO experiment. It also ensures the required level of precision on the $\bar{\nu_e}$ energy reconstruction, a  Requirements of this electronics and main test results are presented hereafter.

\section{Electronics concept}
\label{sec:Electronics concept}

\begin{figure}[h]
	\begin{center}
		\includegraphics[width=0.3\linewidth]{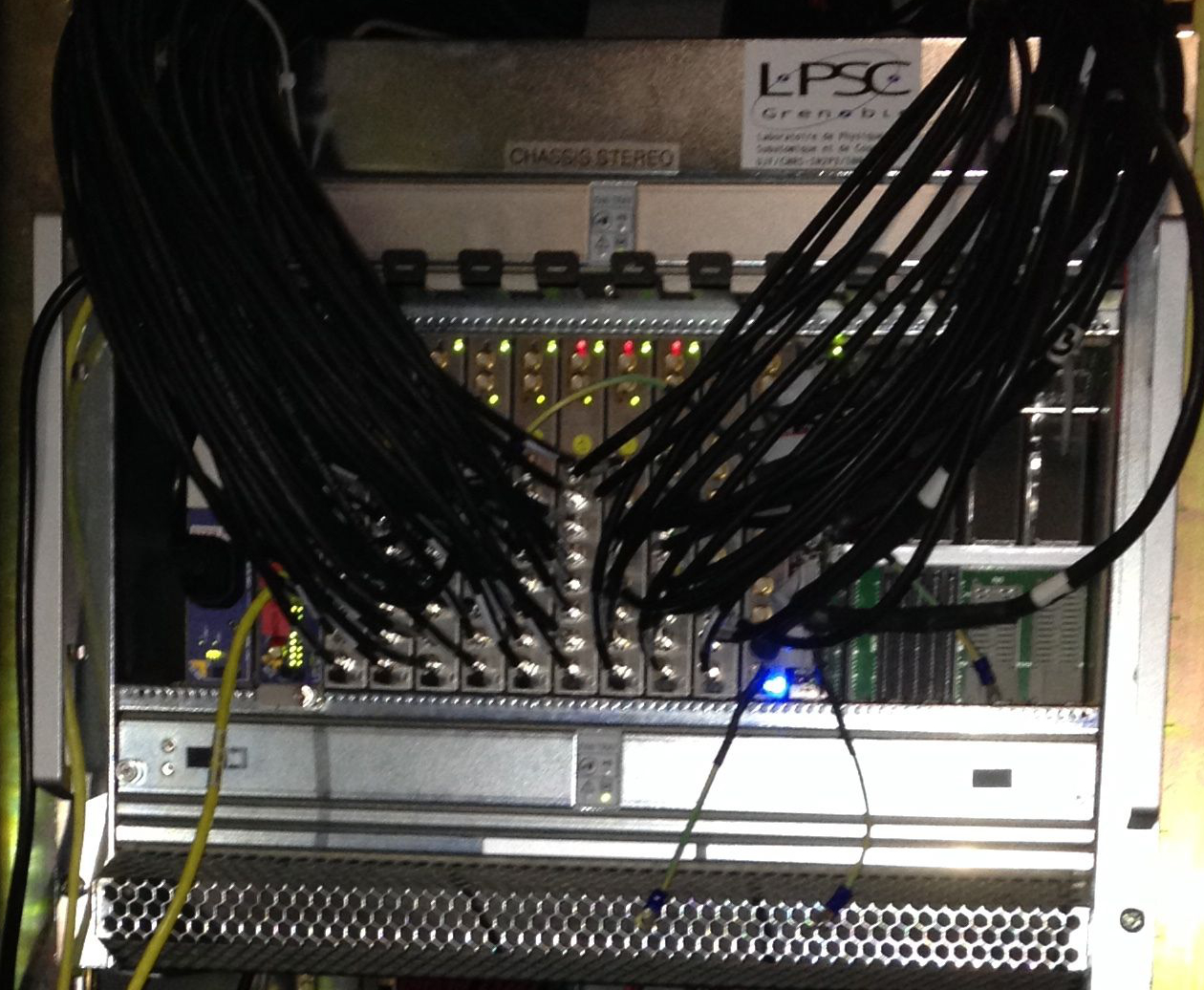}\hspace*{1cm}
		\includegraphics[width=0.35\linewidth]{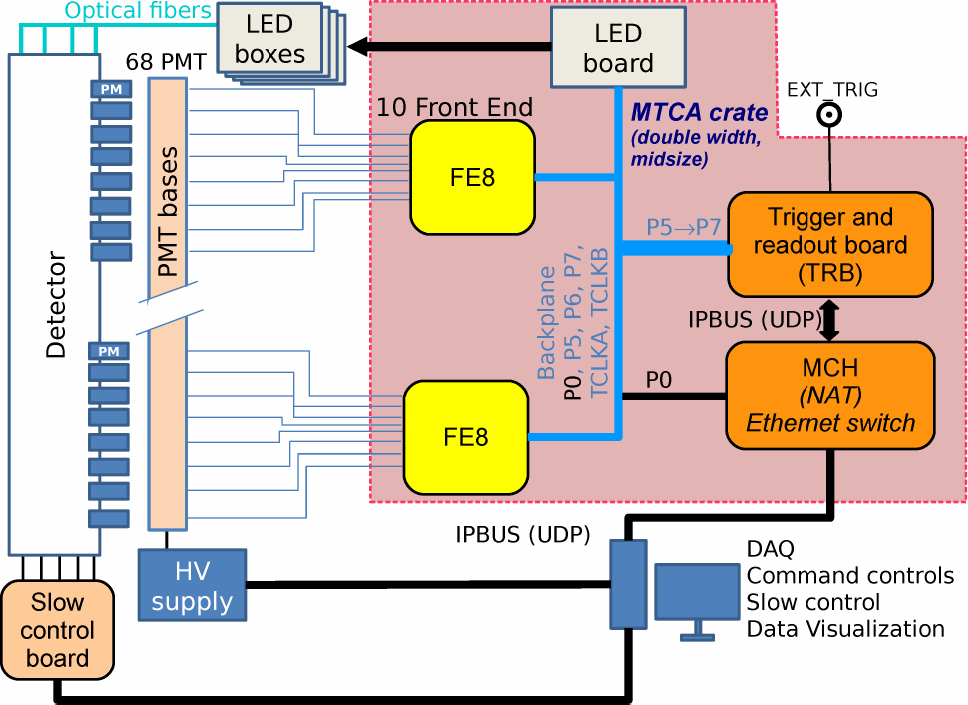}
		\caption{Left panel: picture of the STEREO electronics hosted in the \micro TCA crate. Right panel: scheme describing the working principle of the STEREO electronics.}
		\label{fig:electronics_description}
	\end{center}
\end{figure}

The STEREO electronics (see \cite{Bourion2016} for a full description) are hosted in a \micro TCA crate, for compactness, modularity and versatility (see Fig\;\ref{fig:electronics_description}). It performs triggering in two stages with various selectable conditions, processing, readout and on-line calibration of the 68 photo-multipliers (PMTs) -- 48 for the detector and 20 for the veto -- whose 100 to 200\,ns signals are continuously digitised at 250\,MSamples/s.
For that purpose the \micro TCA crate is equipped with 10 front-end boards (FE8). One trigger and readout board (TRB) is used to perform the second level trigger (T2), to collect and aggregate the processed data provided by FE8. Additionally, for detector performance monitoring, on-line calibration by LED synchronised with the data acquisition is allowed. These electronics have been designed to handle at least 1\,kHz mean trigger rate.
In order to perform a precise measurement of the $\bar{\nu_e}$ energy, the pedestal is corrected using a high-pass IIR. In addition, the ADC non-linearity has been measured using LEDs to be less than 1\,\% up to 1500\,photo-electrons (PE), corresponding to 8\,MeV $\bar{\nu_e}$.

\begin{figure}[h]
	\begin{center}
		\includegraphics[width=0.4\linewidth]{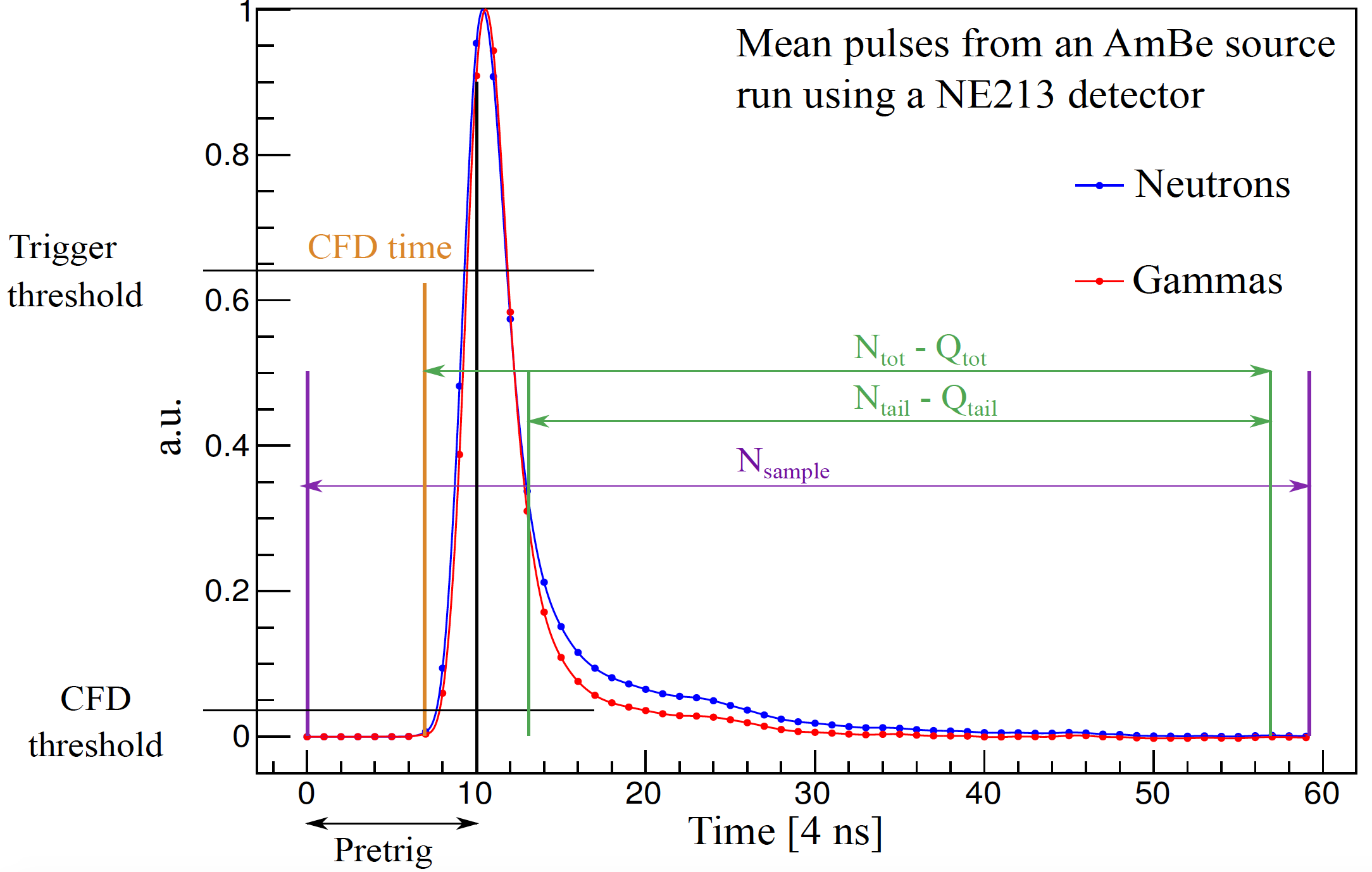}\hspace*{1cm}
		\includegraphics[width=0.3\linewidth]{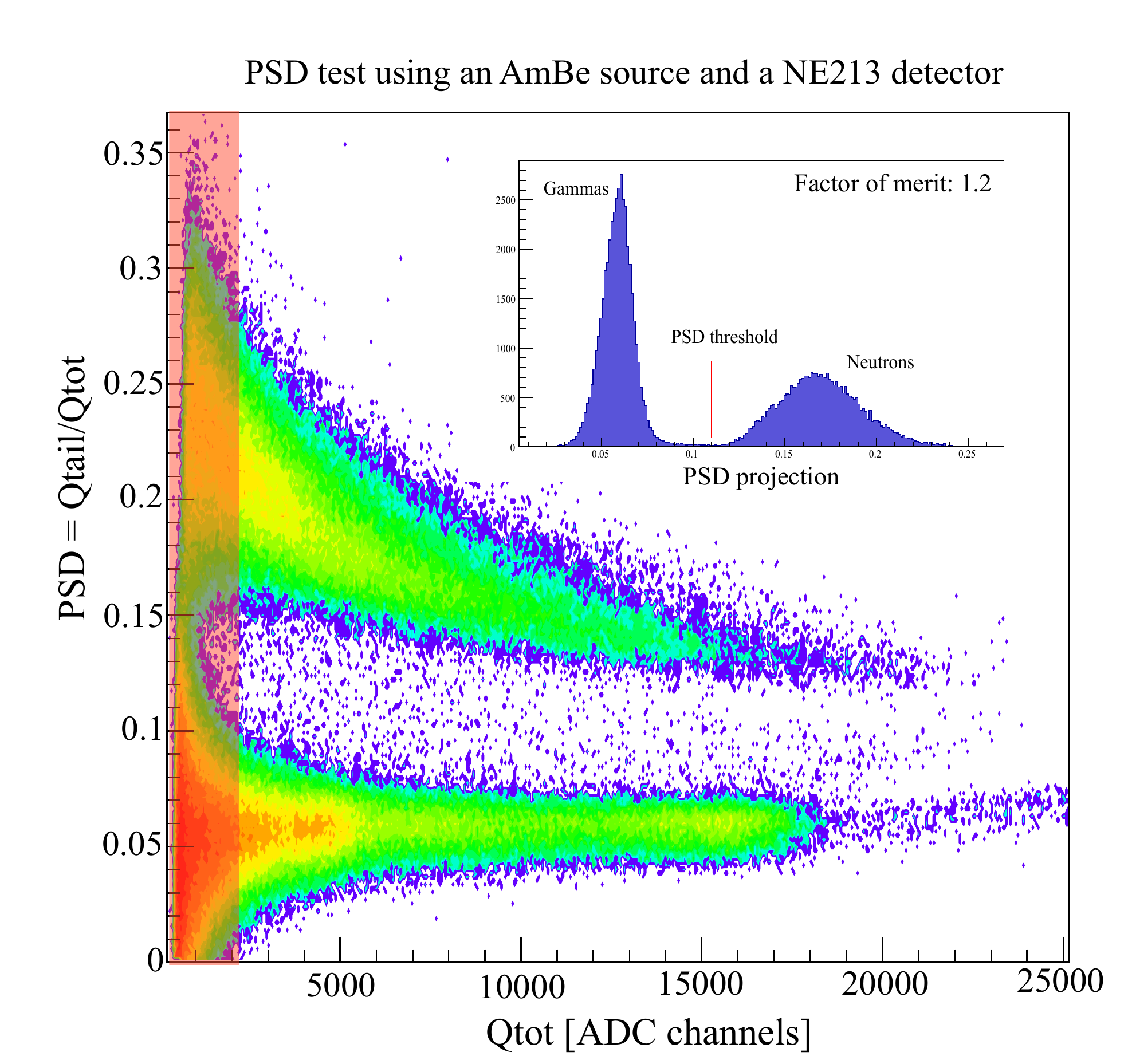}
		\caption{Left panel: scheme of signal processing on FE8. Right panel: PSD using a NE213 detector combined to STEREO electronics to detect neutrons and $\gamma$ from an AmBe source.}
		\label{fig:signal_processing}
	\end{center}
\end{figure}

Both single PE calibration and physics run can be performed thanks to FE8 pre-amplifiers two selectable gains (respectively $\times$20 and $\times$1). Then, signals pass through the ADC controlled and readout by an FPGA (flashed for easy acquisition changes). The Constant Fraction Discriminator is computed on boards, to find the pulse start and perform the PSD by computing the total pulse charge and the charge of the end of the signal only, as depicted in Fig\;\ref{fig:signal_processing}. Using the optional debug mode, samples used for the computation are readout for further signal processing off-line. Using the PSD, it has been shown with an AmBe source and a NE213 detector that a clean fast neutron background rejection (see Fig\;\ref{fig:signal_processing}) is allowed.

A first level threshold (T1) is implemented on each FE8 on amplitude or charge. This trigger can be applied on individual channels or the sum of 4 first/last channels of FE8 or all 8. Trigger conditions can therefore be adapted to the physics case thanks to this flexibility(see \textit{i.e.} Sec.\;\ref{sec:Improvement of muon veto efficiency}). The T1 is sent by the concerned FE8 to the TRB which can then perform a T2 such as on the hits multiplicity or on the total charge in specific parts of the detector.

\section{LED calibrations}

The TRB is also used for on-line calibration using LEDs by driving LED pattern sequences and period as well as sending trigger to the LED board and to the FE8. LED light is injected into the detector via optical fibres. To this aim, 5 LED boxes containing 6 LED each are controlled by the LED board. Different measurements are possible:
\begin{itemize}
\item Single photo-electron calibration (low intensity LED).
\item PMT collection efficiency and liquid scintillator monitoring (high intensity LED).
\item PMT charge linearity measurement (preselected pattern of 6 high intensity LEDs).
\end{itemize}

\section{Improvement of the muon veto efficiency}
\label{sec:Improvement of muon veto efficiency}

\begin{figure}[h]
	\begin{center}
		\includegraphics[width=0.25\linewidth]{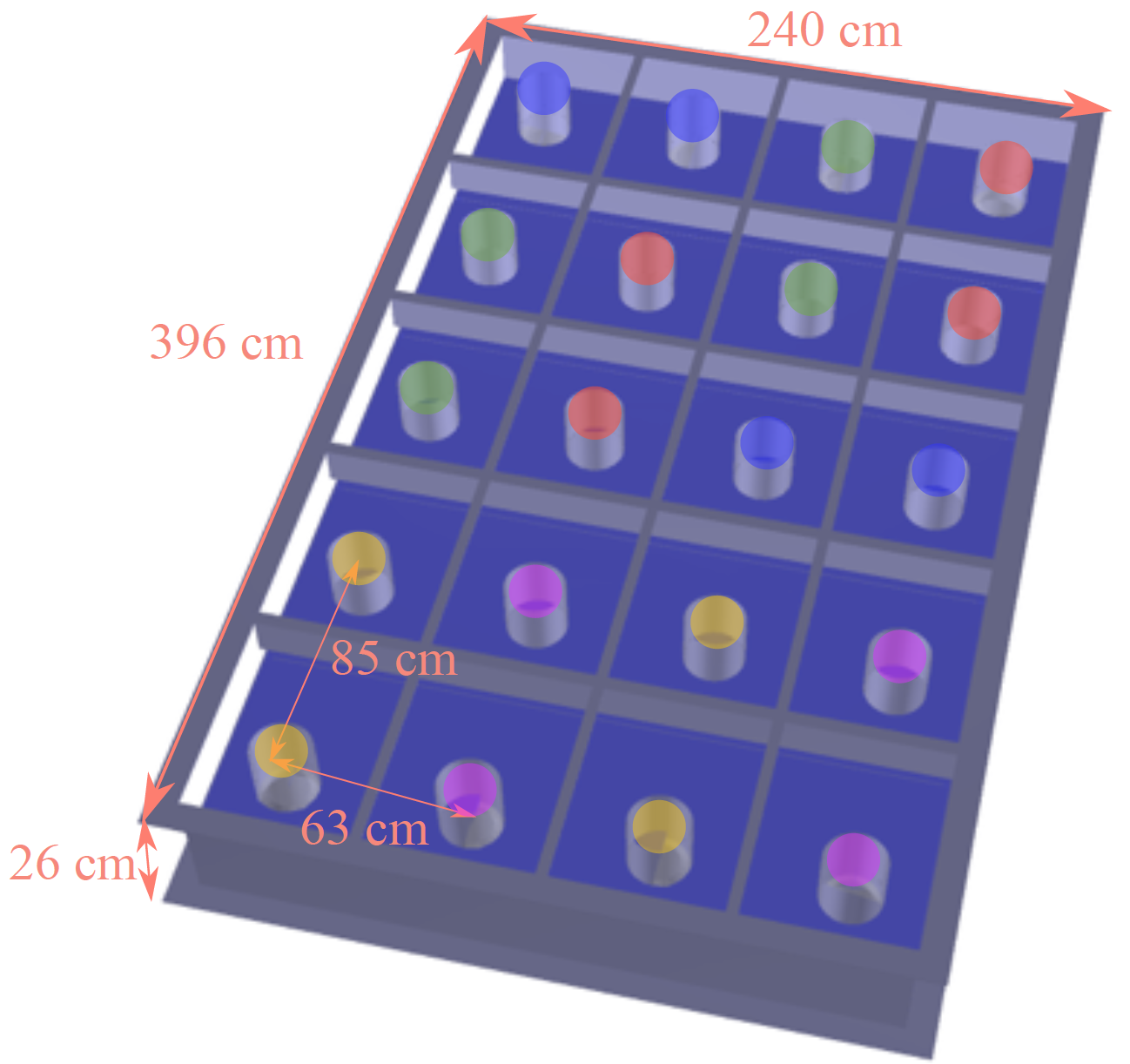}\hspace*{1.5cm}
		\includegraphics[width=0.35\linewidth]{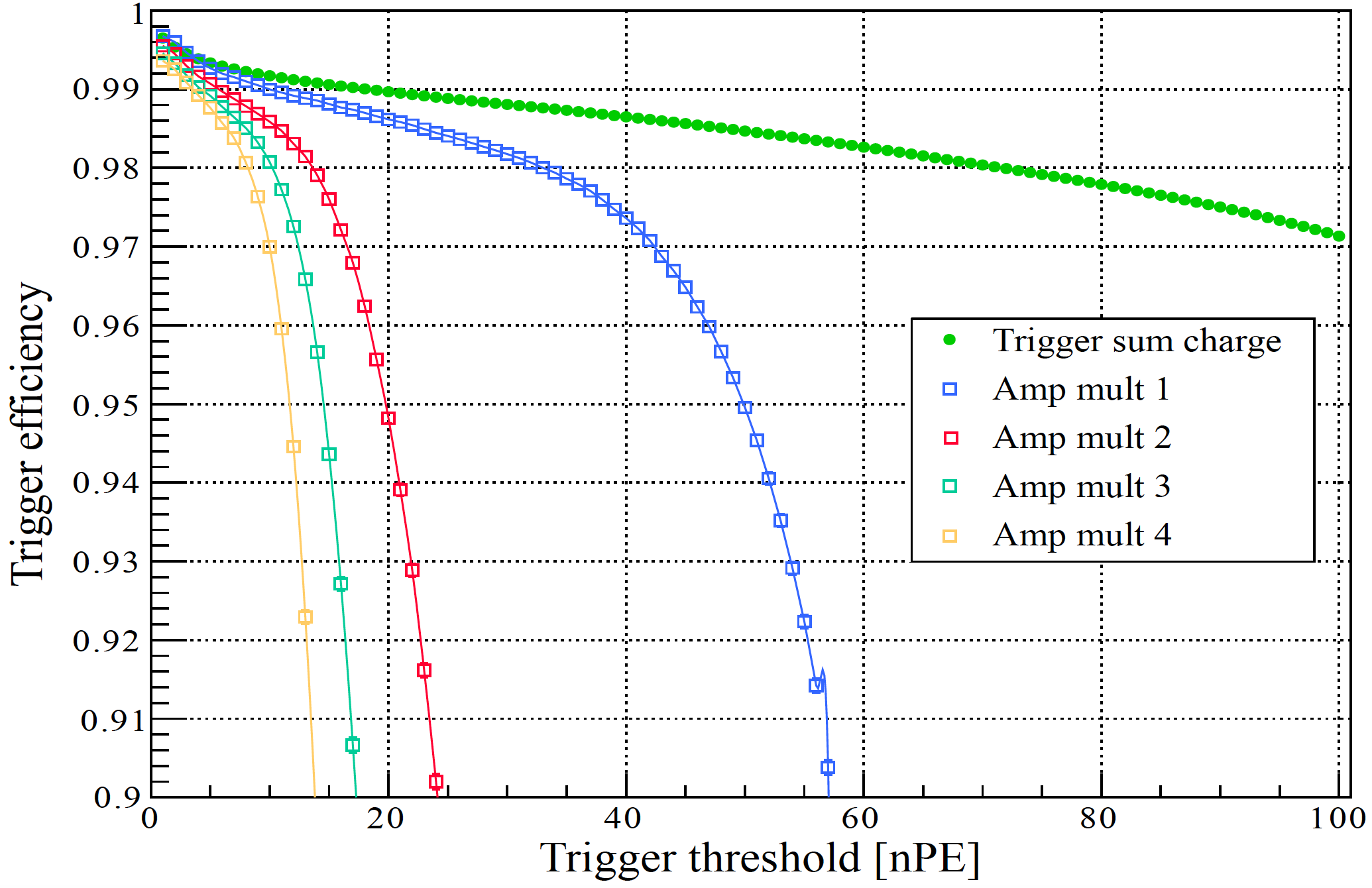}
		\caption{Left panel: muon veto schematic view. The 4-PMTs electronics channels triggering groups are shown by colors. Right panel: muon veto efficiency for different trigger configurations.}
		\label{fig:veto}
	\end{center}
\end{figure}

The muon veto is a water Cherenkov detector equipped with 20 PMTs and used to tag muons (see Fig\;\ref{fig:veto}). This tag is used to apply an off-line veto on the previous muon time window. The device is made of a 2.5\,m$^3$ fiducial volume covering the top of the detector. Thanks to the trigger using the sum of 4-channels blocks, a dedicated mapping is set to increase the veto efficiency and to reduce its sensitivity to the trigger threshold, as shown in Fig\;\ref{fig:veto}. Since the expected muon rate is of about 500\,Hz, this improvement is crucial, reducing drastically background from cosmics. Another advantage of this trigger configuration is to get a reduced sensitivity to $\gamma$ rays and avoid external background triggering. Indeed, Cherenkov light produced by Compton scattering e$^-$ or e$^-$/e$^+$ pairs is localised compared to the muon Cherenkov light.

\section{Conclusion}

Dedicated STEREO electronics have been tested and fulfil all requirements in terms of energy resolution and accuracy for the precise measurement of the $\bar{\nu_e}$ energy spectrum, associated to the on-line LED calibration. Electronics specifications are also adapted to the PSD to reject fast neutron background and the adaptable logic has been used to optimise the muon veto efficiency and accuracy, allowing for a better background rejection.

\end{document}